\documentclass[dvips]{article}

\usepackage{icrc2011}

\title{Cosmic ray spectral hardening due to dispersion of source
injection spectra}

\newcommand{\etal}{\MakeLowercase{\textit{et al. }}} 
\shorttitle{Yuan \etal Spectral hardening of cosmic rays}

\authors{Qiang Yuan$^{1}$, Bing Zhang$^{2}$, Xiao-Jun Bi$^{1}$}
\afiliations{$^1$Key Laboratory of Particle Astrophysics, Institute of 
High Energy Physics, Chinese Academy of Sciences, Beijing 100049, China\\ 
$^2$Department of Physics and Astronomy, University of Nevada Las Vegas,
Las Vegas, NV 89154, USA}
\email{yuanq@ihep.ac.cn, zhang@physics.unlv.edu, bixj@ihep.ac.cn}

\abstract{The cosmic ray (CR) energy spectra measured with ATIC, CREAM 
and PAMELA showed that there is remarkable hardening for rigidity of 
several hundred GV. We propose that this hardening is due to the 
superposition of spectra from a population of sources, e.g., supernova 
remnants (SNRs), whose injection spectral indices have a dispersion. 
Adopting proper model parameters the observational data can be well 
explained. It is interesting that the injection source parameters are 
similar with that derived from gamma-ray observations of SNRs, which 
may support the SNR-origin of CRs. Furthermore this
mechanism provides an alternative explanation of the ``ankle-cutoff''
structure of the ultra high energy CR spectra.
}

\begin{document}
\maketitle

\section{Introduction}

The origins, acceleration processes, propagation and interactions of
cosmic rays (CRs) are still open questions, even after about one century
since the discovery of CRs. It is generally believed that Galactic
CRs (GCRs) are accelerated by the astrophysical shocks like supernova
remnants. After the production, GCRs are then injected into the
interstellar space and propagate diffusively in the Galactic magnetic
field. During the propagation, CRs will interact with the interstellar
medium, radiation field and magnetic field, which result in spallation
and energy loss of the particles, and the production of secondary
particles (see \cite{2007ARNPS..57..285S} for a review).

In recent years more and more accurate data of the CR spectra and
composition are available. The balloon-borne experiment Advanced Thin
Ionization Calorimeter (ATIC) measured the CR spectra of various
species of nuclei and showed the deviation from single power-laws
of the spectra \cite{2007BRASP..71..494P,2009BRASP..73..564P}.
ATIC data also revealed the difference between proton and Helium
spectra. Another balloon experiment Cosmic Ray Energetics And Mass
(CREAM) measured the energy spectra of the major species from proton
to iron with relatively high precision, and reported a remarkable
hardening of the spectra of most heavy species at $\sim200$ GeV/nucleon
\cite{2009ApJ...707..593A,2010ApJ...714L..89A}. Such spectral
features were confirmed most recently by the satellite experiment
Payload for Antimatter Matter Exploration and Light-nuclei Astrophysics
(PAMELA). PAMELA measured the proton and Helium spectra up to rigidity
$1.2$ TV with high precision, and the data show clearly a spectral
hardening at rigidity $\sim 200$ GV \cite{2011Sci...332...69A}, which
is basically consistent with the results of CREAM and ATIC. The spectral
indices of proton and Helium were also found significantly different.

The hardening of the CR spectra challenges the conventional idea that
the spectra below the so-called ``knee'' are simple power-laws. More
complicated scenarios of CR origin, acceleration or propagation are
needed to explain the data. Models possibly to explain such a spectral
hardening include the multi-component sources \cite{2006A&A...458....1Z},
or the nonlinear particle acceleration scenarios where the feedback of 
CRs on the shock is essential (e.g., \cite{1999ApJ...526..385B,
2001RPPh...64..429M,2007ApJ...661L.175B,2010ApJ...718...31P}).

In \cite{2011PhRvD..84d3002Y} we propose that the hardening of the
observed CR spectra is due to the dispersion of the injection spectra 
of a source population such as supernova remnants (SNRs). The basic
fact that the superposition of a series of spectra with different 
power-law indices would result in an asymptotic hardening of the final 
spectra of CRs was recognized long ago \cite{1972ApJ...174..253B}.
Such an effect was also employed to explain the ``GeV excess'' of 
Galactic diffuse $\gamma$-rays observed by EGRET 
\cite{1998ApJ...507..327P,2001A&A...377.1056B}. 

The observations of various kinds of candidate high energy CR sources
such as SNRs, active galactic nuclei and gamma-ray bursts, show indeed
there is significant dispersion of the source spectra. For example, the 
spectral modeling of the $\gamma$-ray emission from several SNRs observed 
by Fermi and ground-based Cherenkov telescopes show that the low and 
high energy spectral indices are $\gamma_1\approx 2.15\pm0.33$ and
$\gamma_2\approx 2.54\pm0.44$, assuming a hadronic scenario of the
$\gamma$-ray emission \cite{2011APh....35...33Y}. In the following
we will show how such a scenario can natually explain the observational
data of the CR spectra.

\section{Model and result}

We assume that GCRs are originated from SNR-like sources. The injection 
spectrum of each source is assumed to be a broken power-law function of 
rigidity. The spectral indices are assumed to be Gaussian distributed 
around the average values. The break rigidity $R_b$ is shown to be several 
to tens of GV \cite{2011APh....35...33Y}. Here we assume the logarithm of 
$R_b$ is randomly distributed in some range. The normalization of each 
source is derived assuming the same total energy of CRs above 1 GeV for 
all sources. The detailed parameters are given below, according to the 
fit to the observational CR data.

The CR propagation is calculated with the public
GALPROP\footnote{http://galprop.stanford.edu/} code 
\cite{1998ApJ...509..212S}. In this work we adopt the diffusive 
reacceleration frame of the propagation model, with the main parameters 
$D_0=5.8\times10^{28}$ cm$^2$ s$^{-1}$, $\delta=0.33$, $v_A=32$ km 
s$^{-1}$ and $z_h=4$ kpc. The input source spectra are adopted as the 
superposed ones of many sources. The B/C ratio is found to be well 
consistent with data and insensitive to the source spectrum.

For the low energy spectra (below $\sim 30$ GV), the force-field 
approximation is adopted to model the solar modulation effect 
\cite{1968ApJ...154.1011G}. The modulation potential $\Phi$ varies
from $\sim 200$ MV to $\sim 1400$ MV, depending on the solar activity.
The modulation potential was estimated to be $450-550$ MV for the
time when the proton and Helium data used in this work were recorded 
by PAMELA \cite{2011Sci...332...69A}. In this work we adopt $\Phi=550$ MV
for proton and Helium. For Carbon, Oxygen and Iron nuclei the low
energy data used are from HEAO3, for which we adopt $\Phi=750$ MV.
A higher modulation potential for HEAO3 was also found in 
\cite{2011ApJ...729..106T}.

We further adopt a break/cutoff at high rigidities ($\sim$PV) to 
approach the ``knee'' of the CR spectra. The physical models of 
the knee include the propagation/leakage effect from the Galaxy, or 
interactions with background particles \cite{2004APh....21..241H}.
Phenomenologically we adopt two kinds of cutoff/break to model the 
knee structure of the total spectra: a sub-exponential cutoff case 
with the energy spectrum above the injection break 
$R^{-\gamma_2}\exp\left[-(R/R_c)^{1/2}\right]$, and a broken power-law 
case with energy spectrum above the injection break $R^{-\gamma_2} 
(1+R/R_c)^{-1}$. In both cases we assume the break rigidities $R_c$ of 
different nuclei are the same, i.e., the break energies are $Z$-dependent. 
The parameter $R_c$ is adopted to be a constant instead of that varying 
for different sources. This assumption is reasonable for the 
propagation/leakage and interaction models. However, we may note that 
for the acceleration limit models, the break energies of different 
sources should have a dispersion. It is tested that the results with 
dispersion of the break rigidities are similar to that with a proper 
constant $R_c$.

\begin{figure*}[!htb]
\centering
\includegraphics[width=0.65\columnwidth]{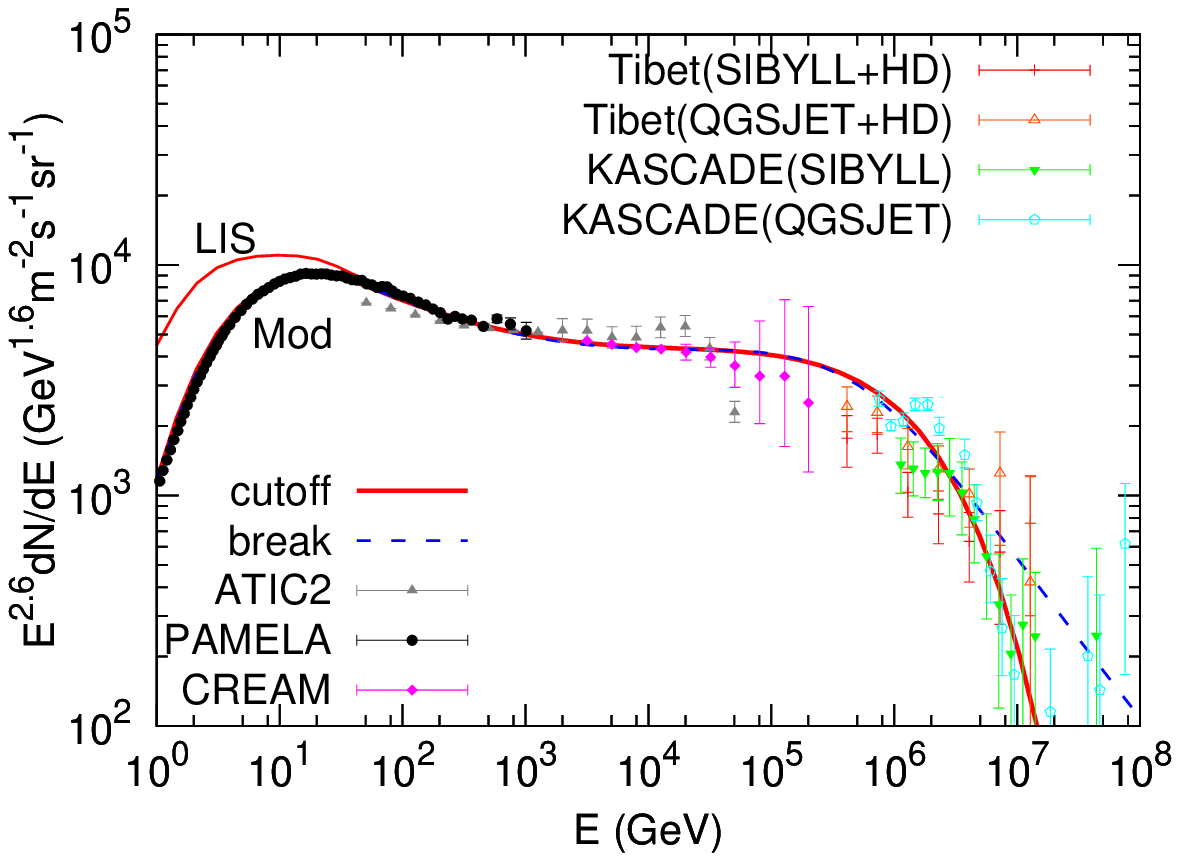}
\includegraphics[width=0.65\columnwidth]{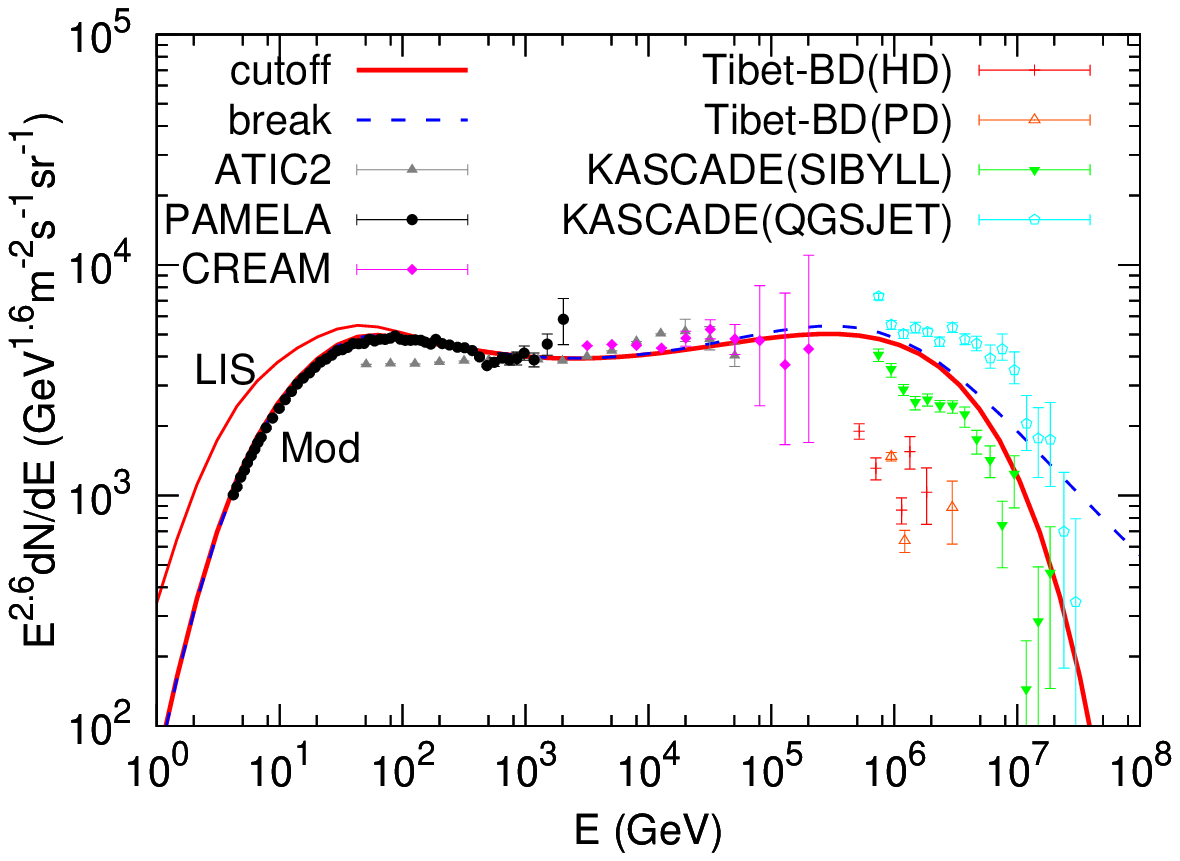}
\includegraphics[width=0.65\columnwidth]{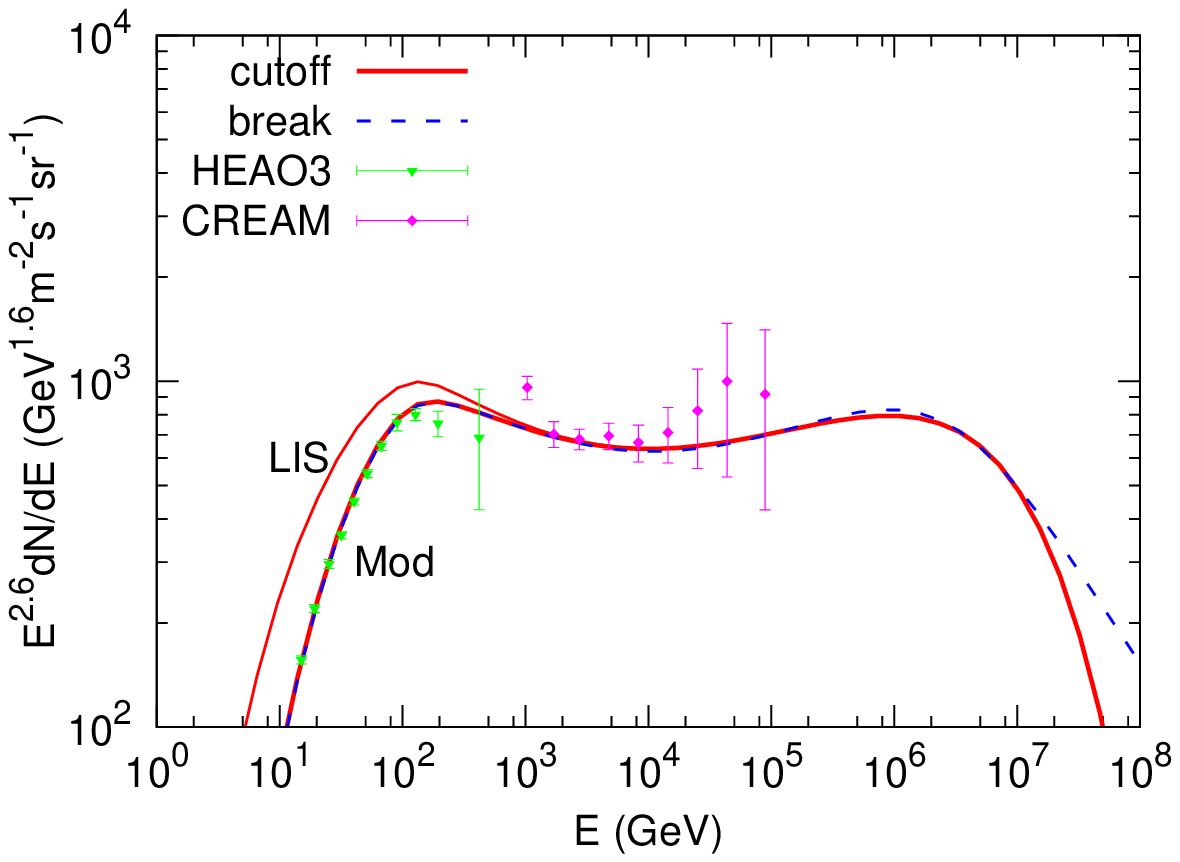}
\includegraphics[width=0.65\columnwidth]{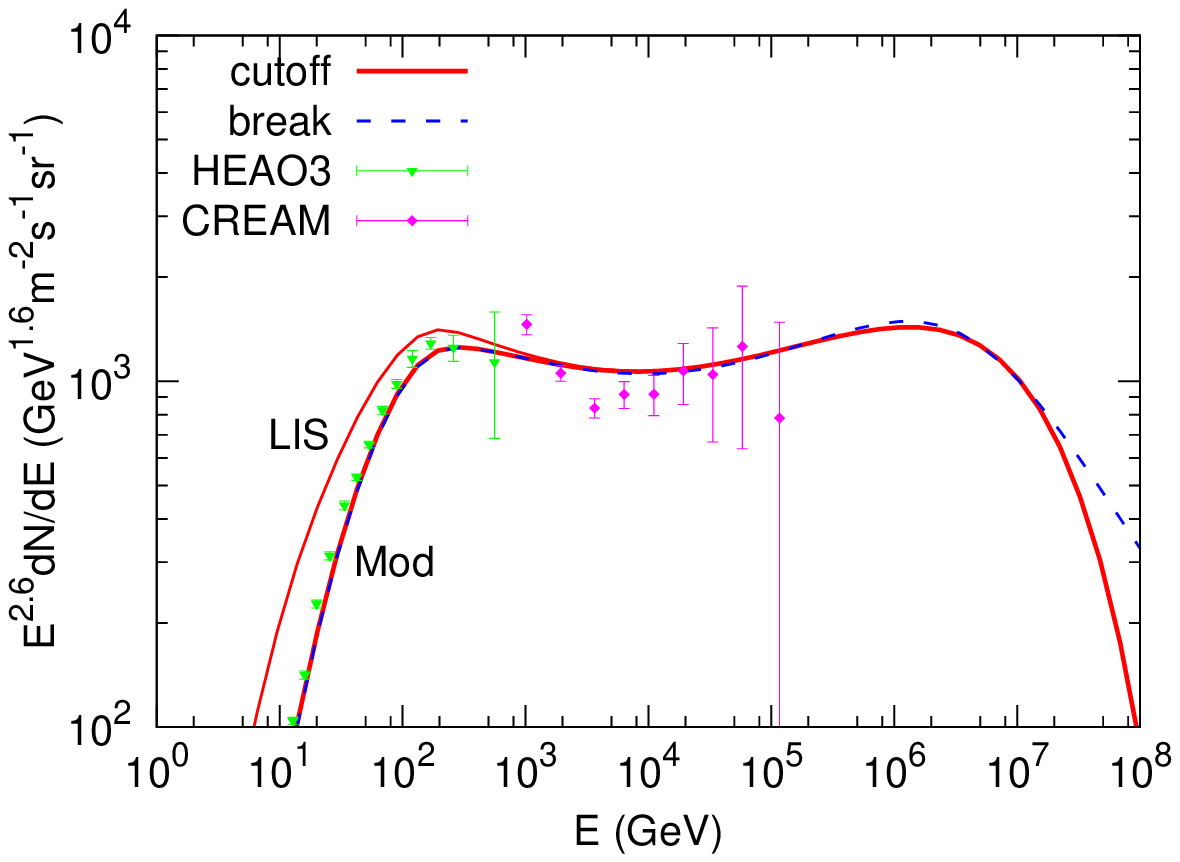}
\includegraphics[width=0.65\columnwidth]{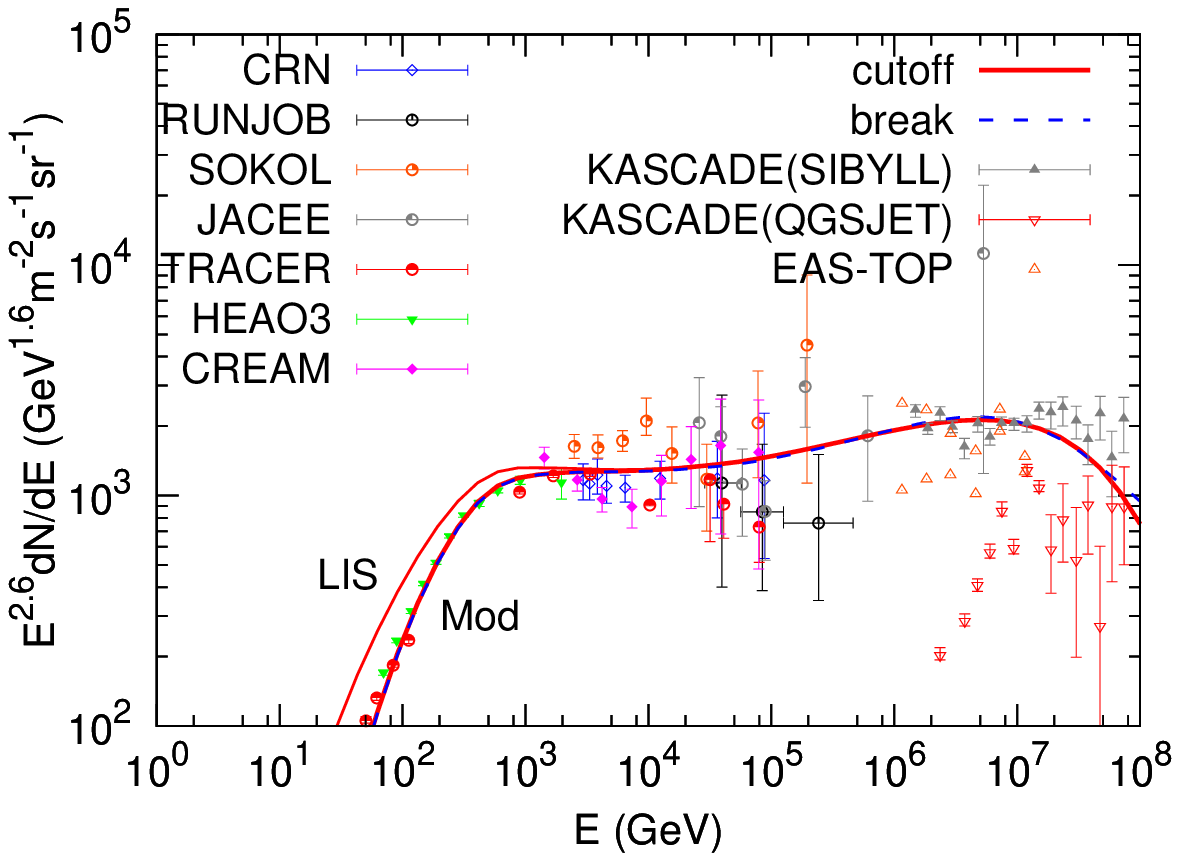}
\includegraphics[width=0.65\columnwidth]{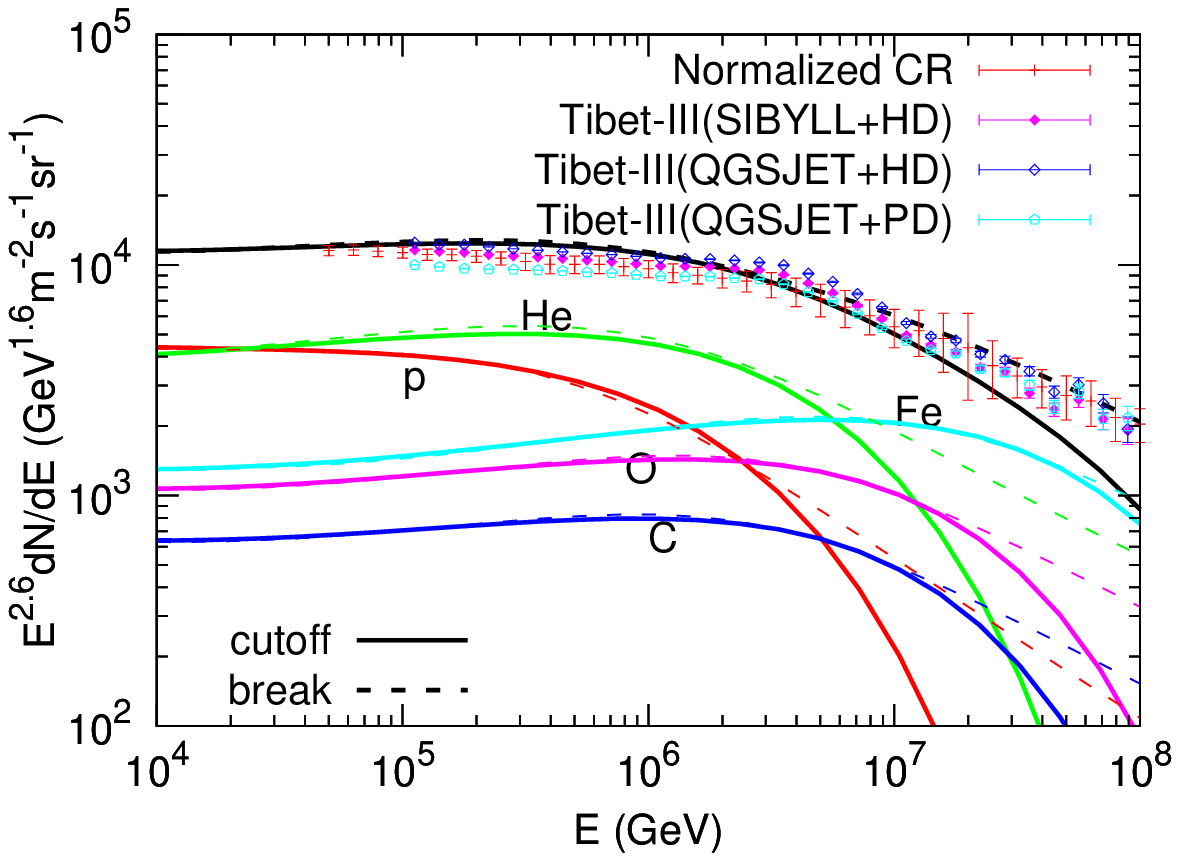}
\caption{Energy spectra of proton (top-left), Helium (top-middle),
Carbon (top-right), Oxygen (bottom-left), Iron (bottom-middle) and the 
all-particle one (bottom-right). The solid line in each panel represents 
a sub-exponential cutoff behavior of the high energy spectra around the 
knee region, while the dashed line is for broken power-law type.
For references of the data please refer to \cite{2011PhRvD..84d3002Y}.
}
\label{fig:f1}
\end{figure*}

\begin{table*}[htb]
\centering
\caption{Source parameters: injection spectra $\gamma_1$, $\gamma_2$ and
break rigidity $R_b$, high energy cutoff rigidity $R_c$ and solar modulation
potential $\Phi$.}
\begin{tabular}{ccccccc}
\hline \hline
  & & $\gamma_1$ & $\gamma_2$ & $R_b$ & $R_c$ & $\Phi$ \vspace{-0mm} \\
  & &            &            & (GV)  & (PV)  & (GV) \\
\hline
         & p      & $1.95\pm0.20$ & $2.52\pm0.28$ & $[5,30]$ & $0.5$ & $0.55$ \\
  cutoff & He     & $1.95\pm0.20$ & $2.50\pm0.33$ & $[5,30]$ & $0.5$ & $0.55$ \\
         & C,O,Fe & $1.95\pm0.20$ & $2.58\pm0.35$ & $[5,30]$ & $0.5$ & $0.75$ \\
  \hline
         & p      & $1.95\pm0.20$ & $2.52\pm0.25$ & $[5,30]$ & $0.5$ & $0.55$ \\
  break  & He     & $1.95\pm0.20$ & $2.50\pm0.30$ & $[5,30]$ & $0.5$ & $0.55$ \\
         & C,O,Fe & $1.95\pm0.20$ & $2.58\pm0.32$ & $[5,30]$ & $0.5$ & $0.75$ \\
  \hline
  \hline
\end{tabular}
\label{table1}
\end{table*}

The calculated energy spectra of proton, Helium, Carbon, Oxygen, Iron
and the total spectrum, together with the observational data are shown
in Figure \ref{fig:f1}. The parameters used in the calculation are compiled
in Table \ref{table1}. We find relatively good agreement between the 
model calculation and the observational data. It is interesting that
the injection source parameters are similar to those inferred from the 
$\gamma$-ray observations of SNRs. This might be evidence that SNRs are 
the sources of GCRs below $\sim$PV.

Note that our model prediction is a gradual hardening of the CR energy 
spectra, which seems to be consistent with the overall structures of data 
in a wide energy range. However, if we focus on the detailed structures 
revealed by individual experiment, we may still find some inconsistency. 
For example the PAMELA data show a very sharp dip at $\sim200$GV and a 
gradual softening below the break rigidity \cite{2011Sci...332...69A}.
The CREAM data indicate that all species of nuclei experience a
hardening at $\sim200$ GeV/n \cite{2010ApJ...714L..89A}. The very fast 
break of the spectra shown in both PAMELA and CREAM data cannot be
well reproduced in the present model. Before giving a conclusive
judgement about this issue, we may need future better measurements
of wide energy band spectra by e.g., the Alpha Magnetic Spectrometer 
(AMS02\footnote{http://ams.cern.ch/}) and the Large High Altitude Air 
Shower Observatory (LHAASO, \cite{2010ChPhC..34..249C}).

\section{Impact on secondary particles}

\begin{figure}[!htb]
\centering
\includegraphics[width=\columnwidth]{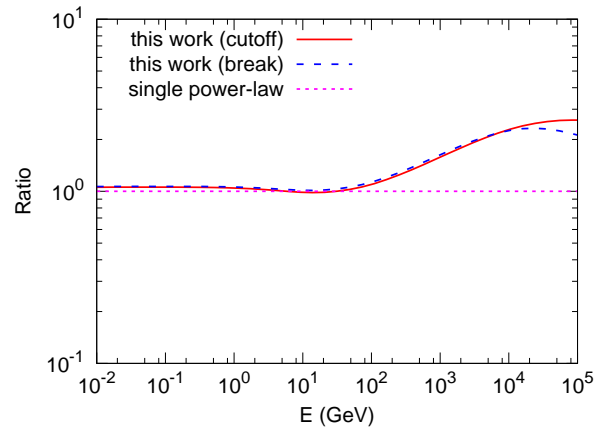}
\caption{Ratio of the hadronic component of the diffuse $\gamma$-rays
between the dispersion scenario expectation and the single power-law model.
}
\label{fig:f2}
\end{figure}

The hardening of the primary CR particles should have imprints on the
secondary particles such as positrons \cite{2011MNRAS.414..985L}, 
diffuse $\gamma$-rays, and antiprotons \cite{2011PhRvD..83b3014D}.
In \cite{2011arXiv1108.1023V} a systematic study of the secondary 
particles including B/C, antiprotons and diffuse $\gamma$-rays
in different scenarios of the spectral hardening was performed.
As an example, here we calculate the predicted hadronic-origin 
diffuse $\gamma$-ray fluxes in this scenario. The $\gamma$-ray
production spectrum from $pp$ inelastic interactions is calculated
using the parameterization given in \cite{2006ApJ...647..692K}.
The contribution to $\gamma$-rays from heavy nuclei in both the
projectile and target particles is approximated with a nuclear 
enhancement factor $\epsilon_M=1.84$ \cite{2009APh....31..341M}.
We show in Figure \ref{fig:f2} the ratios of the hadronic component
of the diffuse $\gamma$-ray fluxes between our model expectation 
and that of the traditional single power-law CR spectrum. It is 
shown that the $\gamma$-ray flux will also experience a hardening 
above $\sim 50$ GeV. The most remarkable hardening effect will be in 
the very high energy range ($>$TeV), which is out of the
capability of Fermi telescope.

Since the total diffuse $\gamma$-rays consist with hadronic, leptonic 
and the extra-galactic components, the hadronic one is not a direct 
observable. As we know that in the Galactic plane the diffuse 
$\gamma$-rays should be dominated by the hadronic component 
\cite{2004ApJ...613..962S}, we expect such a predication can be tested 
with high precision diffuse TeV $\gamma$-ray observation of the Galactic 
plane.

For positrons and antiprotons we will expect similar results, however,
the quantitative effects should depend on the source distribution and
propagation model. The current antiproton data from PAMELA can not
probe such a hardening effect yet. Also the PAMELA positron excess
above several GeV should not be originated from the hardening of the
CR proton and Helium spectra.

\section{Alternative explanation of the ultra high energy CR spectra?}

We also note that the dip-cutoff structure shown in Figure \ref{fig:f1} 
is very similar to the ankle and Greisen-Zatsepin-Kuzmin (GZK) cutoff
structure of the ultra high energy CRs (UHECRs). If the UHECRs are 
originated from a population of sources instead of a major one, it is 
expected that the UHECR spectra should also suffer from such a hardening 
due to the superposition effect. It might provide us an alternative
point of view to understand the observational energy spectra of UHECRs.

\begin{figure}[!htb]
\centering
\includegraphics[width=\columnwidth]{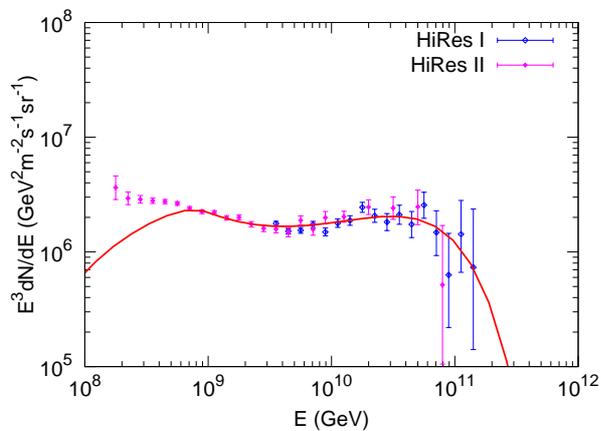}
\caption{Calculated energy spectra of UHECRs for pure protons,
compared with the HiRes data \cite{2008PhRvL.100j1101A}.
}\label{fig:f3}
\end{figure}

We show an illustration of the predicted UHECR spectra, together with
the HiRes data \cite{2008PhRvL.100j1101A}, in Figure \ref{fig:f3}.
Here the chemical composition of UHECRs is assumed to be pure protons.
The injection spectrum of UHECRs is adopted to be a broken power-law 
function with an exponential cutoff \cite{2006PhRvD..74d3005B}. The 
logarithm of break energy $\log(E_b/{\rm eV})$ is assumed to be
uniformly distributed in $[17,18]$, and the spectral index is 
$2.0\pm 0.2$ below $E_b$ and $3.6\pm 0.6$ above $E_b$. The cutoff energy 
$E_c$ is adopted to be $5\times 10^{19}$ eV. The result shows a good
description of the observational data.

If the UHECRs are produced at cosmological distances, however, the 
interactions between UHECRs and the cosmic background photons
are unavoidable \cite{2006PhRvD..74d3005B}. Therefore if the superposition
effect is responsible for the shape of the UHECR spectrum, we may in turn 
expect that UHECRs are produced locally or even in the Galaxy 
\cite{2010PhRvL.105i1101C}. More generally, it is possible that both the
superposition and the interaction effects are in operation to give
the ankle-GZK structure of UHECRs.

\section{Summary}

In summary we propose that the superposition of the energy spectra
from CR sources with a dispersion of injection spectra is responsible 
for the recently reported hardening of the CR spectra. If the CRs are 
indeed originated from a population of sources instead of a single major 
source (e.g., \cite{1997JPhG...23..979E}), such an asymptotic hardening 
effect due to dispersion of source properties is inevitable. It is 
interesting to note that the source parameters derived are similar
to that inferred from $\gamma$-ray observations of SNRs, which might
support the SNR-origin of GCRs below the knee. Finally we discuss the 
same mechanism as an alternative explanation of the ankle-GZK structure 
of UHECR spectra.

This work is supported by NSF under grant AST-0908362, NASA under grants
NNX10AP53G and NNX10AD48G, and National Natural Sciences Foundation of 
China under grant 11075169, and the 973 project under grant 2010CB833000.

\clearpage

\end{document}